\documentclass[conference]{IEEEtran}
\IEEEoverridecommandlockouts

\usepackage{cite}
\usepackage{amsmath,amssymb,amsfonts}
\usepackage{algorithmic}
\usepackage{graphicx}
\usepackage{textcomp}
\usepackage{xcolor}
\usepackage{hyperref}
\def\BibTeX{{\rm B\kern-.05em{\sc i\kern-.025em b}\kern-.08em
    T\kern-.1667em\lower.7ex\hbox{E}\kern-.125emX}}
\begin{document}

\title{COMPLETE RECONSTRUCTION OF THE TONGUE CONTOUR THROUGH ACOUSTIC TO ARTICULATORY INVERSION USING REAL-TIME MRI DATA\\}

\author{\IEEEauthorblockN{Sofiane Azzouz\textsuperscript{1}, Pierre-André Vuissoz\textsuperscript{2}, Yves Laprie\textsuperscript{1}}
\IEEEauthorblockA{\textit{\textsuperscript{1}Université de Lorraine, CNRS, Inria, F-54000 Nancy, France}\\
\textit{\textsuperscript{2}Université de Lorraine, Inserm, IADI U1254, F-54000 Nancy, France}\\
sofiane.azzouz@loria.fr, pa.vuissoz@chru-nancy.fr, yves.laprie@loria.fr}
}

\maketitle

\begin{abstract}
Acoustic articulatory inversion is a major processing challenge, with a wide range of applications from speech synthesis to feedback systems for language learning and rehabilitation.

In recent years, deep learning methods have been applied to the inversion of less than a dozen geometrical positions corresponding to sensors glued to easily accessible articulators. It is therefore impossible to know the shape of the whole tongue from root to tip. In this work, we use high-quality real-time MRI data to track the contour of the tongue. The data used to drive the inversion are therefore the unstructured speech signal and the tongue contours.  
Several architectures relying on a Bi-LSTM including or not an autoencoder to reduce the dimensionality of the latent space, using or not the phonetic segmentation have been explored.
The results show that the tongue contour can be recovered with a median accuracy of 2.21 mm (or 1.37 pixel) taking a context of 1 MFCC frame (static, delta and double-delta cepstral features).
\end{abstract}


\section{Introduction}
Articulatory-acoustic inversion (A-to-A) aims to recover the underlying articulatory parameters from the acoustic speech signal. It has evolved over time, adopting various methods to overcome the challenges of indeterminacy since many vocal tract shapes can give the same speech spectrum \cite{Atal78}. The first approaches consisted of exploiting articulatory synthesis \cite{rubin1981articulatory} simulating speech production to analyze speech and recover the shape of the vocal tract over time. 
Several solutions have been explored either by using an articulatory model \cite{10.1121/1.3514544}
as a geometrical constraint or by using variational calculus to directly obtain the area function approximated by tubes \cite{schoentgen1995direct}.

However, the mismatch between the underlying articulatory synthesizer and the analyzed speaker limits the performance of such an approach, and machine learning based techniques exploiting either electromagnetic articulography (EMA) \cite{Uria2012DeepAF} or X-ray microbeam data \cite{Westbury2005} have been adopted. The advantage is to connect the input speech signal with the geometrical position of sensors without any intermediate explicit production model.

Subsequently, the modeling of gesture dynamics has itself been inspired by articulatory phonology, as in  \cite{deng1994statistical} devoted to automatic speech recognition by using atomic speech units based on overlapping articulatory features.  Then, a statistical approach  of inversion was developed \cite{toda2004acoustic} based on Gaussian Mixture Models (GMM) to model articulatory distributions from complex acoustic signals. Simultaneously, \cite{hiroya2004estimation} employed a speech production model based on hidden Markov models (HMM), combining HMM with articulatory dynamics to improve inversion accuracy.

More recently, the emergence of deep learning techniques, particularly derived from recurrent neural networks (RNN), has delivered significant results in A-to-A inversion most often by using Mel-Frequency Cepstral Coefficients (MFCCs) as input. Bi-LSTM was used in \cite{liu2015deep,parrot2020independent}, CNN-BiLSTM in \cite{shahrebabaki2020sequence}, Bi-GRNN in \cite{wu2023speaker}, and TCN in \cite{siriwardena2023secret}. To achieve better results, \cite{wang2023two} additionally incorporated phonetic segmentation as input. The issue with this last method is that phonetic segmentation is always required for inference. To address this problem while leveraging the additional information provided by phonetic transcriptions, multitask learning frameworks have been proposed, such as in \cite{siriwardena2022acoustic}, where phonemic labels are jointly predicted alongside articulatory variables as targets.

However, the greatest limitation of these approaches lies in the nature of EMA data, since it concerns the position of a few sensors glued to easily accessible articulators (the lips, the central lower incisor and the front part of the tongue in the oral cavity).
 This is a major limitation that reduces the value of A-to-A inversion, since the entire lower part of the tongue, the pharynx and the larynx, whose position has a direct impact on the length of the vocal tract, remain inaccessible.
Current approaches are therefore unable to reconstruct the complete geometry of the vocal tract, and especially the tongue contour, which greatly reduces their usefulness. 
For this reason, real time Magnetic Resonance Imaging (rt-MRI) with the partially denoised acoustic signal was used  to train the recovery of midsagittal images of the vocal tract from the input speech signal represented as Mel-Generalized Cepstral Line Spectral Pairs (MGC-LSP) \cite{csapo2020speakerInterSpeech}.
 In his work Csapo showed that LSTMs are the most efficient for this task. However, the images  used have low resolution (68x68 pixels), the quality of the denoised signal is quite poor, and the result is an MRI image that suffers from both the image characteristics (low resolution and some MRI artefacts) and the inaccuracies of the inversion, making them very difficult to exploit.

The work we present here shows that it is possible to recover the contour of the whole tongue by using, during training, the contours obtained from rt-MRI images of good resolution (136x136 pixels) and a good quality denoised speech signal.

\section{Dataset}

\subsection{Corpus}\label{AA}
The corpus includes recordings of a French-speaking female, consisting of 2100 sentences, totaling approximately 3.5 hours of recordings. This corpus was recorded at the Centre Hospitalier Régional Universitaire de Nancy and contains 178 acquisitions, each with a duration of 80 seconds and comprising 4000 images. The corresponding audio recordings have a sampling rate of 16 kHz. The rt-MRI images were captured with a resolution of 136 x 136 pixels and a frame rate of 50 (fps). The software Astali \cite{fohr2015importance} was used to perform forced alignment of the speech with the transcriptions and
to obtain phonetic segmentation. The phonetic annotations were then carefully manually corrected by an expert.
We randomly divided our dataset by acquisitions into 80\% for training, 10\% for validation, and 10\% for testing.

\subsection{Data preprocessing}
From the audio signal, MFCCs were calculated along with their first (delta 1) and second (delta 2) derivatives, using a fixed number of coefficients set to 13. The calculation was performed with a window of 25 milliseconds and a MFCC frame has been computed every 10ms.

Following \cite{parrot2020independent}  we incorporated a context window of 11 frames, consisting of 5 previous frames, 5 subsequent frames, and the current frame, to enrich the temporal information, thus covering a total temporal window of 125 ms (without considering the initial and final $\Delta$ and $\Delta\Delta$).
For tongue contours we used the automatic tracking approach proposed based on a Mask R-CNN proposed in \cite{ribeiro2024automatic}, as visualized in Figure 1, which gives very good results on our MRI data.
Each tongue contour includes 50 X and Y coordinate points.
For phonetic segmentation, we initially encoded the segmentations in one-hot format. Silent segments between sentences (given by the segmentation) were discarded because no relevant articulatory information can be recovered. Indeed, it is not possible to consider that the tongue is always in a neutral resting position during silences. This may involve breathing, swallowing or other gestures that do not correspond to the production of a speech signal. Silences in the middle of a sentence, on the other hand, are maintained. Following this, the MFCCs and the tongue contours were normalized as described in [10]. We normalized the MFCCs and the tongue contours by subtracting the mean and dividing by the standard deviation. For the tongue contours, we used the mean and standard deviation calculated from the 30 preceding and following recordings to ensure a consistent distribution of the extracted features.

An MFCC frame is calculated every 10 ms, while each MRI image covers 20 ms. To align the MRI images with the MFCC frames, an intermediate contour was obtained by interpolating the tongue contours of two successive images.

\begin{figure}[ht] 
  \centering
  \includegraphics[width=0.5\textwidth]{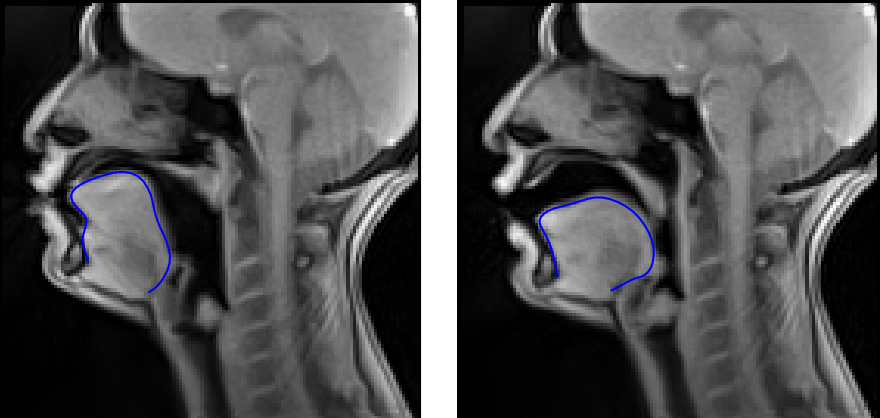} 
  \caption{Tongue contour (from the tongue root to the sublingual cavity) tracked in two images of the rt-MRI film} 
  \label{fig:original_contours} 
\end{figure}

\section{Methods}
\subsection{Model architecture}
We designed a Bi-LSTM model for the simultaneous prediction of articulatory contours and phonemes from an audio signal, as illustrated in Figure 2. We were inspired by the model presented in \cite{parrot2020independent}.

The model starts with a dense layer of 300 units, followed by two bidirectional LSTM layers, each with 300 units. It takes as input a feature vector of dimension 429 (corresponding to one frame of 11 MFCC vectors) and then passes through two additional dense layers, each with 300 units.
For the single-task model, the output is generated by a dense layer, producing 100 points representing the tongue contours (50 points for X and Y coordinates). When using an autoencoder, the output corresponds to a 16-dimensional latent space vector, from which the autoencoder reconstructs the 100 points of the tongue contour.
In the multi-task configuration, the model provides two outputs: one identical to that of the single-task model and the other, obtained through an additional dense layer, which provides classification probabilities for the 43 phonemes of French.

\begin{figure*}[ht]
    \centering
    \includegraphics[width=0.6\textwidth]{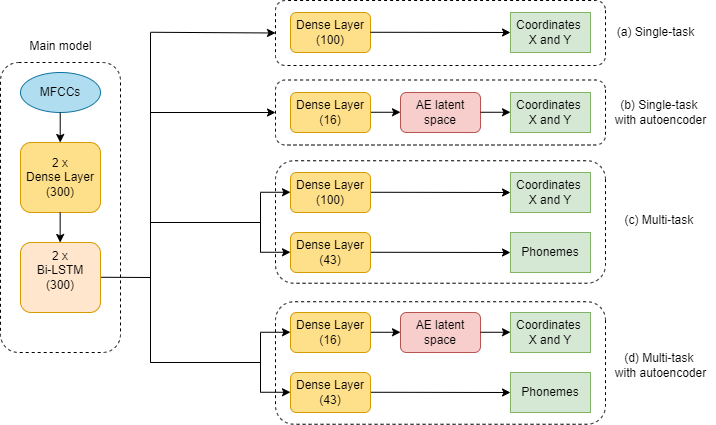} 
    \caption{Single-task (a, b) and multi-task (c, d) architectures with (b, d) and without (a, c) autoencoder}
    \label{fig:grande_image}
\end{figure*}
\subsection{Loss function}

Acoustic inversion is a regression task where the most commonly used loss function is the Mean Squared Error (MSE). This function aims to minimize the squared differences between the actual and predicted trajectories. When a classification task, such as phoneme recognition, is also included, the Cross Entropy loss function is employed to evaluate the divergence between the predicted probability distribution for each phoneme and the actual phoneme labels. 

\begin{equation}
\text{Cross Entropy} = - \sum_{i=1}^{n} \sum_{c=1}^{C} y_{i,c} \log(\hat{y}_{i,c})\label{eq1}
\end{equation} 

where \(n\) represents the total number of observations in the dataset, \(C\) is the total number of classes, \(y_{i,c}\) and \(\hat{y}_{i,c}\) represent the true and predicted probabilities, respectively, that example \(i\) belongs to class \(c\).

\begin{equation}
\text{MSE} = \frac{1}{n} \sum_{i=1}^{n} \left( y_i - \hat{y}_i \right)^2\label{eq2}
\end{equation}

\(y_i\) and \(\hat{y}_i\) represent the true and predicted values of the output for example \(i\), respectively.

For predicting both the contours of the tongue and classifying the phonemes, these two loss functions are combined. Weighting is applied using alpha
$(\alpha = 1)$
variable to adjust the relative importance of Cross Entropy, thereby balancing the impact of each task on the overall model training.

\begin{equation}
L(y_i, \hat{y}_i) = \text{MSE}(y_i, \hat{y}_i) + \alpha \cdot \text{CrossEntropy}(y_i, \hat{y}_i)\label{eq3}
\end{equation}

\subsection{Evaluation of the model}
For evaluating the model, we used several metrics on the test data. We measured the unnormalized root mean square error (RMSE) and the median, both in millimeters, for single-task and multi-task models. Additionally, for the multi-task models, we also evaluated phonetic accuracy, which was calculated using cross-entropy. These metrics allowed us to compare the performance of the models from different perspectives.

\begin{equation}
\text{RMSE} = \sqrt{\frac{1}{n} \sum_{i=1}^{n} \left( y_i - \hat{y}_i \right)^2}\label{eq4}
\end{equation}

\subsection{Experiments}
We carried out two distinct experiments: the first involves predicting only the tongue contours, while the second includes both the prediction of tongue contours and phonetic segmentation. For each experiment, we assessed performance both with and without the use of an autoencoder. We used the autoencoder  designed to estimate the shape of the tongue contour during continuous speech \cite{ribeiro2022autoencoder}. Instead of predicting the complete shape directly, our model learns to predict the autoencoder's representation. The predicted parameters are then used to generate the tongue contour using the autoencoder's decoder

This comparison will allow us to evaluate the impact of the autoencoder on the effectiveness of tongue contour prediction and the accuracy of phonetic segmentation.
Additionally, we performed four further experiments using only the single-task model by varying the context window (1, 3, 5, and 7 frames) to investigate the role of temporal information in the model’s performance.

\subsection{Model Parameters}
Our model was trained for 300 epochs with a batch size of 10. The Adam optimizer was used with an initial learning rate of 0.001. Early stopping with a patience of 5 epochs was applied on the validation data, resulting in the training being stopped if no improvement was observed. The entire code was implemented using PyTorch.

\section{Results}
Table 1 presents the performance of the different models evaluated for the prediction tasks. The models are classified as follows: Single-task Context Window 1 (ST-1), Multi-task with Autoencoder (MT-AE), Single-task Context Window 7 (ST-7), Multi-task (MT), Single-task with Autoencoder (ST-AE), Single-task Context Window 5 (ST-5) Single-task Context Window 3 (ST-3) and Single-task (ST). All models converged, except for ST-1, which only converged after 401 training epochs.

The results reveal that ST-1 achieved the best performance, with a median value of 2.21 mm and an RMSE of 2.52 mm, making it the top-performing model in both metrics. Following closely, MT-AE ranked second, with a median of 2.28 mm, an RMSE of 2.58 mm, and the highest phoneme prediction accuracy at 75.54\%, while MT reached 64.45\%. Both ST-7 and MT share a median value of 2.31 mm, although MT exhibits a slightly higher RMSE of 2.63 mm compared to ST-7's 2.61 mm. Models such as ST-AE and ST-5 performed similarly, with a median of 2.33 mm and RMSE values of 2.63 mm. Finally, ST and ST-3 showed the least favorable results, with medians of 2.36 mm and 2.34 mm, respectively, and slightly higher RMSE values of 2.64 mm and 2.65 mm, placing them among the lower-performing models.

\begin{table}[htbp]

\caption{rmse, median and accuracy phonemes (ACC) in all experiments. ST: Single-Task, MT: Multi-Task, ST-AE: Single-Task with AutoEncoder, MT-AE: Multi-Task with AutoEncoder, ST-1: Single-Task with 1 context window, ST-3: Single-Task with 3 context windows, ST-5: Single-Task with 5 context windows, ST-7: Single-Task with 7 context windows.}
\begin{center}
\resizebox{0.95\linewidth}{!}{%
\begin{tabular}{|c|c|c|c|}
\hline
\textbf{Model} & \textbf{RMSE(mm)} & \textbf{MEDIAN(mm)} & \textbf{ACC(\%)}\\ \hline
        ST & 2.64 $\pm$\, 1.27 &2.36 & -\\\hline
        MT & 2.63 $\pm$\, 1.37 & 2.31 & 64.45\\ \hline
        ST-AE & 2.63 $\pm$\, 1.31 & 2.33  & -\\ \hline
        MT-AE & 2.58 $\pm$\, 1.32 & 2,28 & \textbf{75.54}\\ \hline
        ST-1 & \textbf{2.52} $\pm$\, 1.31 & \textbf{2.21}  & -\\ \hline
        ST-3 & 2.65 $\pm$\, 1.32 & 2.34  & -\\ \hline
        ST-5 & 2.63 $\pm$\, 1.32 & 2.33  & -\\ \hline
        ST-7 & 2.61 $\pm$\, 1.31 & 2.31  & -\\ \hline
\end{tabular}
}
\label{tab1}
\end{center}
\end{table}


\section{Discussion}

The results show non-significant variations among the evaluated models for the prediction tasks. Models with the same context window sizes (11) but different architectures (MT, ST-AE, MT-AE) perform better than the ST model. Among these, the MT-AE model, which integrates phonetic prediction and utilizes an autoencoder, stands out with the best median score of 2.28 mm and an RMSE of 2.58 mm. Additionally, MT-AE achieves the highest phoneme prediction accuracy at 75.54\%. This suggests that incorporating phonetic segmentations and using autoencoders can positively influence performance, although the differences are not substantial. Models with different context window sizes but the same architecture (ST-1, ST-3, ST-5, and ST-7) also show better performance compared to the ST model. The ST-1 model delivers the best median score of 2.21 mm and an RMSE of 2.52 mm, surpassing all other models, including those with different architectures. This indicates that context window size positively affects performance, although the variations are not significant enough to have a decisive impact.

We also noted that the model struggles to predict rapid tongue movements and variations smaller than the window size, as illustrated by the right image in Figure 3, which shows a significant difference between the original and predicted contours, with an RMSE of 4.61 mm. In contrast, the left image demonstrates a better prediction of the tongue contour, with an RMSE of 2.49 mm. The pauses between sentences have been removed, but there are still quite long pauses (often containing breaths) within sentences, which are the origin of the largest deviations. 

 Additionally, although the quality of contour segmentation is very good, it is not perfect, and the model cannot surpass the contours provided by segmentation. Compared to the literature, our results are moderately good, but our approach stands out by enabling the prediction of the entire tongue contour, unlike existing methods.

\begin{figure}[ht] 
  \centering
  \includegraphics[width=0.5\textwidth]{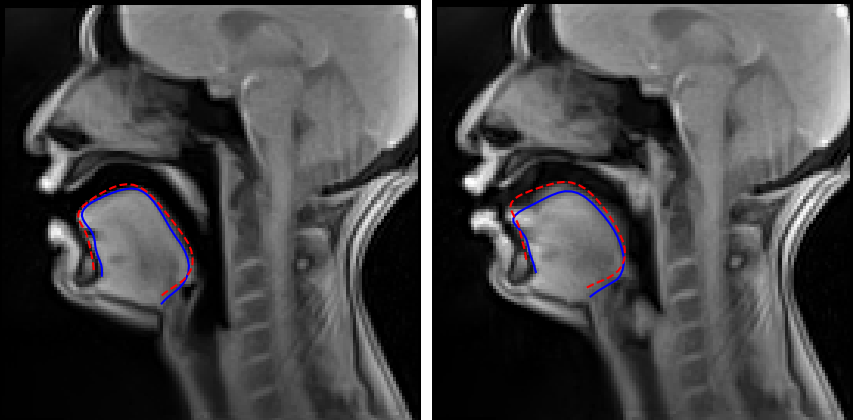} 
  \caption{Example of two large inversion deviations with respect to the expected contours tracked in rt-MRI film. On the left, an image corresponding to a respiration, and on the right, an image acquired at the time just before the closure for a dental stop. The red dotted line represents the predicted contours, the solid blue line gives  the original contours.   } 
  \label{fig:infrence_contours} 
\end{figure}

The video\footnote{\url{https://youtu.be/v9CbKJ0ZmzI}} shows the tongue contours inverted from the signal displayed in blue on the images. 
\section{Conclusion}

In this work the complete tongue contour was successfully recovered from the speech signal recorded at the same time as rt-MRI images from which the contours were derived. 
The advantage of using tongue contours instead of raw MRI images is to spare an additional post-processing step from an inverse image, whose contours are likely to be less marked and more difficult to identify. This makes it possible to exploit the inversion results for a wide range of potential applications.

These results demonstrate for the first time the possibility to recover complete tongue contours form speech signal with less than 2.21 mm median RMSE to the corresponding segmentation of the real time MRI images.

We will continue this work for all the contours of the vocal tract. We decided to concentrate on the tongue first, since it is the most mobile and deformable articulator, and also the one that plays the most important phonetic and acoustic role. 

Although we consider the results of automatic tracking to be excellent in the vast majority of images, there are still a few small errors in the vicinity of the tongue tip. We are therefore also working on improving tracking, and another possibility would be to jointly invert the contour and the raw image by modifying the loss function.

However, there is still a further step to be taken to move from the speech recorded in the MRI machine to natural speech, i.e. without the Lombard effect due to the intense noise of the machine, or the effect of lying down, even if the speech recorded in the MRI machine has been denoised very effectively.

\clearpage

\bibliographystyle{IEEEtran}
\bibliography{main.bbl}

\end{document}